\documentclass{kapproc}
\usepackage{procps} 
\usepackage{graphicx}
\upperandlowercase
\normallatexbib

\begin{document}

\articletitle[Optical and magneto-optical properties of moderately
correlated systems]{Ab initio calculations of the optical and
magneto-optical properties of \\ moderately correlated systems:\\ accounting for correlation
effects}

\author{A. Perlov$^1$, S. Chadov$^1$, H. Ebert$^1$, L. Chioncel$^2$, A.
Lichtenstein$^2$, M. Katsnelson$^3$}

\begin{center}
\affil{$^1$University of Munich, Butenandstrasse 5-13, D-81377, Munich, Germany \\
$^2$University of Nijmegen, NL-6526 ED Nijmegen, The Netherlands \\
$^3$Uppsala University, P.O.Box 530, S-751 21 Uppsala, Sweden}
\end{center}
\begin{abstract}
The influence of dynamical correlation effects on the magneto-optical
properties of ferromagnetic Fe and Ni has been investigated. In addition
the temperature dependence of the self-energy and its influence on the DOS and
optical conductivity is considered. Magneto-optical properties
were calculated on the basis of the one-particle Green's function, which
was obtained from the DMFT-SPTF procedure. It is shown that dynamical
correlations play a rather important role in weakly correlated Fe and
substantially change the spectra for moderately correlated
Ni. Magneto-optical properties obtained for both systems are found in
better agreement with experiment than by conventional LDA calculations.

\end{abstract}

\section{Introduction}

Much  information on  the  electronic structure  of  magnetic solids  is
gained by  optical and magneto-optical measurements,  being useful tools
for  analyzing  the  dispersion  of  (quasi-particle)  bands.   However,
measured optical  and magneto-optical spectra can  hardly be interpreted
without accompanying  theoretical calculations. For this  purpose one in
general  has to solve  a corresponding  many-electron problem,  which is
impossible  without the use of more  or less  severe approximations. For materials
where the  kinetic energy  of the electrons  is more important  than the
Coulomb interactions, the most successful first principles method is the
Local (Spin-)  Density Approximation (L(S)DA) to  the Density Functional
theory (DFT) \cite{HK64},  where the many-body problem is  mapped onto a
non-interacting   system   with   a  one-electron   exchange-correlation
potential approximated by that of the homogeneous electron gas.  For the
last  two  decades {\em  ab  initio}  calculations  of the  optical  and
magneto-optical properties of solids based on this approximation yielded
a good basis for such an interpretation, often leading to a quantitative
agreement between  theoretical and experimental  spectra.  The situation
is very  different when we consider more  strongly correlated materials,
(systems containing f and d electrons) since in all the calculations the
LDA  eigen-energies are  implicitly interpreted  to be  the one-particle
excitation energies of  the system. It is well known  that there are two
possible sources of error connected with that approach: Firstly, the LDA
provides    only   an   approximate    expression   for    the   (local)
exchange-correlation   potential.   Secondly,   even  with   the   exact
exchange-correlation  potential at hand,  one is  left with  the problem
that   there  is   no   known  correspondence   between  the   Kohn-Sham
eigen-energies     and    the    one-particle     excitation    energies
\cite{JG89,KV83,AvB85,Bor85}.

For an                exact description of  the excitation  energies the
non-local self-energy has to be considered. This, however, constitutes a
many-body problem. Therefore,  DFT-LDA calculations must be supplemented
by  many-body  methods to  arrive  at  a  realistic description  of  the
one-particle excitations in correlated  systems. To give an example, let
us mention  the GW approximation  \cite{Hed65} which is well  suited for
the case  of insulators  and semi-conductors and has also  been applied
successfully to transition metals \cite{Hed65,Ary92,RKP93,AG95}. Another
approach    is   to   consider the Hubbard-type   models    where   those
Coulomb-interaction terms are included explicitly that are assumed to be
treated   insufficiently   within   DFT-LDA.    Already   the   simplest
Hartree-Fock  like   realization  of  such  an   approach  called  LDA+U
\cite{AZA91} scheme  allowed to improve considerably  the description of
the optical  and magneto-optical spectra of  strongly correlated systems
(mostly containing rare  earths elements \cite{YOP+96,AAH+99}).  The main
advantage  of  the  LDA+U  scheme  is the  energy  independence  of  the
self-energy  which allows to  use only  slightly modified  standard band
structure methods  for calculating optical  and magneto-optical spectra.
On  the other  hand  the scheme  works  rather good  only for  extremely
correlated systems, where  Coulomb interactions (U) prevail considerably
over  the  kinetic  energy  (bandwidth W).   For  moderately  correlated
systems (U$\approx$W) which applies for  most $3d$ and $5f$ elements and
their  compounds one has to take  into account a non-Hermitian energy dependent
self-energy to get  a  reasonable  description  of the  electronic
structure. Nowadays  there are  several approaches available to deal with this
situation.   The most advanced one is the Dynamical Mean-Field  Theory (DMFT)
\cite{GKKR96}.   DMFT  is a successful  approach  to investigate  strongly
correlated  systems with local  Coulomb interactions.  It uses  the band
structure results calculated, for  example, within LDA approximation, as
input and then missing electronic correlations are introduced by mapping
the  lattice problem  onto an effective single-site  problem  which is
equivalent  to  an Anderson  impurity  model  \cite{A84}.   Due to  this
equivalence  a variety  of approximative  techniques have  been  used to
solve  the   DMFT  equations,  such  as   Iterated  Perturbation  Theory
(IPT) \cite{GKKR96,GK92},    Non-Crossing     Approximation    (NCA)
\cite{KK70,BCW87},  numerical   techniques  like  Quantum   Monte  Carlo
simulations (QMC) \cite{HF86}, Exact Diagonalization (ED)
\cite{GK92,CK94}, Numerical Renormalization Group (NRG)  \cite{B00}, or Fluctuation
Exchange (FLEX) \cite{BS89,KL99,KL02}. The DMFT maps lattice models onto quantum
impurity models subject to a self-consistency condition in such a way
that the many-body problem for the crystal splits into a single-particle
impurity problem and a many-body problem of an effective atom. In fact,
the DMFT, due to numerical and analytical techniques developed to solve
the effective impurity problem \cite{GKKR96}, is a very efficient and
extensively used approximation for energy-dependent self energy
$\Sigma(\omega)$.  At present LDA+DMFT is  the only  available $ab$
$initio$  computational technique which is able to treat  correlated
electronic  systems   close  to  a  Mott-Hubbard  MIT 
(Metal-Insulator Transition), heavy fermions and $f$-electron systems.

Concerning the  calculation of the optical  spectra we have  to face the
following problem: one particle wave  functions are not defined any more
and   the  formalism   has   to  applied  in   the  Green   function
representation.   Such  a  representation   has  already  been derived
\cite{HE99} and successfully applied for calculations in the framework of
Korringa-Kohn-Rostoker    (KKR)   Green-function    method    for   LSDA
calculations. The only drawback of such an approach is that it is highly
demanding as to both computational resources and computational time.

In  this paper  we propose  a simplified  way to  calculate  optical and
magneto-optical   properties   of   solids   in   the   Green   function
representation   based  on   variational  methods   of   band  structure
calculations.

The paper is organized as following: in section 2 the formalism for 
Green's function calculations of optical and magneto-optical properties
that account for many-body effects through an effective self-energy
is presented. Then, the DMFT-SPTF method for the calculation of the
self-energy is considered. In section 3 the obtained results of our
calculations for Fe and Ni are discussed and compared with experimental
ones. The last section 4 contains the conclusion and an outlook.

\section{Green's function calculations of the conductivity tensor}

Optical properties of solids are conventionally described in terms of
either the dielectric function  or the optical conductivity tensor which 
are connected via the simple relationship:
\begin{eqnarray}
\sigma_{\alpha\beta}(\omega)=-\frac{i\omega}{4\pi}
(\varepsilon_{\alpha\beta}(\omega)-\delta_{\alpha\beta})\,.
\end{eqnarray}
The optical conductivity is connected directly to the other optical properties. For
example, the Kerr rotation $\theta_K(\omega)$ and so-called Kerr ellipticity 
$\varepsilon_K(\omega)$ for small angles and $\mid \varepsilon_{xy}\mid
\ll \mid \varepsilon_{xx}\mid$ can be calculated using the expression \cite{AYPN99}:
\begin{eqnarray}
\theta_K(\omega) + i\varepsilon_K(\omega)\ =\  \frac{-\
\sigma_{xy}(\omega)}{\sigma_{xx}(\omega)\left[1+\frac{4\pi}{\omega}\sigma_{xx}(\omega)\right]^{1/2}}
\ \ .
\end{eqnarray}
The reflectivity coefficient $r$ is given by 
\begin{eqnarray}
r = \frac{(n-1)^2 + k^2}{(n+1)^2 + k^2}
\end{eqnarray}
with $n$ and $k$ being the components of the complex refractive index,
namely refractive and absorptive indices, respectively. They are
connected to the dielectric function via:
\begin{eqnarray}
n + ik= (\varepsilon_{xx} + i\varepsilon_{xy})^{1/2}\ .
\end{eqnarray}

Microscopic calculations of the optical conductivity tensor are
based on the Kubo linear response formalism \cite{Kub58}:
\begin{eqnarray}
\label{eq.kubo}
\sigma_{\alpha\beta}(\omega)=
-\frac{1}{\hbar\omega V} \int_{- \infty }^0 d \tau 
e^{-i(\omega+i\eta)\tau} \left<\ [J_\beta(\tau),J_\alpha(0)]\ \right>
\end{eqnarray}
involving the expectation value of the correlator of the electric
current operator $J_\alpha(\tau)$. In the framework of the quasiparticle
description of the excitation spectra of solids the formula can be
rewritten in the spirit of the Greenwood approach and making use of the
one-particle Green function $G(E)$:

\begin{eqnarray}
\label{eq.condbig}
\!\!\!\!\!\!\! \sigma_{\alpha \beta}(\omega) = \frac{i\hbar}{\pi^2 V}
\int_{-\infty}^{\infty}dE\int_{-\infty}^{\infty}\,dE'
f(E-\mu)f(\mu-E') \nonumber \\ 
\Bigg[\frac{ {\rm Tr}\left\{\hat j_{\alpha}\Im G(E')\hat j_{\beta}\,\Im G(E)\right\}}
{(E'-E+i\eta)(\hbar\omega+E-E'+i\eta)}\ \  + \nonumber \\  
\frac{ {\rm Tr}\left\{\hat j_{\beta}\Im G(E')\hat j_{\alpha}\,\Im G(E)\right\}}
{(E'-E+i\eta)(\hbar\omega+E'-E+i\eta)}\,\,\,\Bigg]
,
\end{eqnarray}
where $\Im G(E)$ stands for the anti-Hermitian part of the Green's
function, $f(E)$ is the Fermi  function and $V$ is the
volume of a sample. Taking the zero temperature limit and  making use of
the analytical properties of the Green's function one can get a
simpler expression for the absorptive (anti-Hermitian)  
part of the conductivity tensor:

\begin{eqnarray}
\label{eq.abscond}
\sigma^{(1)}_{\alpha \beta}(\omega) = \ \frac{1}{\pi\omega} 
\int_{E_F-\omega}^{E_F}\,dE \ 
\mbox{tr}\Bigl[\hat{j}_{\alpha}\Im G(E)\hat{j}_{\beta}\Im
G(E+\hbar\omega)\Bigr] .
\end{eqnarray}

The dispersive part of $\sigma_{\alpha \beta}(\omega)$ is connected to
the absorptive one via a Kramers-Kronig relationship. 

The central quantity entering  expression Eq.(\ref{eq.abscond}) is the
one-particle Green's function defined as a solution of the  equation:

\begin{equation}
\label{eq.se}
[\hat H_0+\hat \Sigma(E)-E]\hat G(E)=\hat I\,,
\end{equation}
where $\hat H_0$ is a one-particle Hamiltonian including the kinetic
energy, the electron-ion Coulomb interaction and the Hartree potential,
while the self-energy  $\hat \Sigma(E)$ describes all static and dynamic
effects of electron-electron exchange and correlations. The L(S)DA introduces the
self-energy as a local, energy independent exchange-correlation 
potential $V_{xc}(r)$. As the introduction of such an additional
potential does not change the properties of  $\hat H_0$ we will
incorporate this potential to  $\hat H_{LDA}$ and subtract this term from the
self-energy operator. This means that the self energy $\Sigma$ used in
the following is meant to describe  exchange and correlation effects not accounted
for within LSDA.

With a choice of the complete basis set $\{|i\rangle\}$ the Green's function
can be represented as:
\begin{eqnarray}
G(E)&=&\sum_{ij}{|i\rangle G_{ij}(E)\langle j|}\ ,\label{lmgf}
\end{eqnarray}
with the Green's matrix $G_{ij}$ being defined as 
\begin{eqnarray}
G_{ij}(E)=\left[\langle i|\hat{H}|j\rangle-E\langle i|j\rangle+\langle
i|\hat{\Sigma}(E)| j\rangle\right]^{-1}.
\end{eqnarray}
Dealing with crystals one can make use of Bloch's theorem when choosing
basic functions $|i_{\bf k}\rangle$. This leads to the ${\bf k}$-dependent Green's function matrix
\begin{eqnarray}
G^{\bf k}_{ij}(E) = [ H_{ij}^{\bf k} - EO_{ij}^{\bf k}+\Sigma_{ij}^{\bf
k}(E)]^{-1}\ .
\end{eqnarray}
Introducing the anti-Hermitian part of the Green's function matrix as 
\begin{eqnarray}
{\cal G}_{ij}^{\bf k}(E) = \frac{i}{2}[G_{ij}^{\bf k}(E)-G_{ji}^{\bf k}(E)]
\end{eqnarray}
and taking into account the above mentioned translational symmetry we
obtain the following expression for the absorptive part of the optical conductivity:
\begin{eqnarray}
\label{kubolmto}
\sigma_{\alpha_\beta}^{abs}=\frac{1}{\pi\omega}
\int_{E_F-\hbar\omega}^{E_F} d E \int d^3 k \sum_{ij}
{\mathcal J} ^\alpha_{ij}({\bf k},E) {\mathcal J}^\beta_{ji}({\bf k},E+\hbar\omega)
\end{eqnarray}
with 
\begin{eqnarray}
{\mathcal J} ^\alpha_{ij}({\bf k},E)= \sum_n{\mathcal G}_{in}^{\bf
k}(E)\langle n_{{\bf k}}|\hat j^\alpha|j_{{\bf k}}\rangle 
\end{eqnarray} 

The efficiency and accuracy of the approach is determined
by the choice of $|i_{\bf k}\rangle$. One of the computationally most efficient
variational methods is the Linear Muffin-Tin Orbitals method \cite{A75} which allows one 
to get a rather accurate description of the valence/conduction band in
the range of about 1~Ry, which is enough for the calculations of the
optical spectra ($\hbar\omega < 6-8$~eV). This method has been used in
the present work. A detailed description of the application of the above
sketched approach in the framework of LMTO can be found elsewhere \cite{PCE03}.

\subsection{Calculation of the self-energy}

The key point for accounting of many-body correlations in the
present approach is the choice of approximation for the self-energy. As
it was discussed in the Introduction one of the most elaborated modern
approximation is DMFT.

For the present work we have chosen one of the most computationally
efficient variants of DMFT: Spin polarized $T$-matrix plus
fluctuation exchange (SPTF) approximation \cite{KL02}, which is based on
the general many-body Hamiltonian in the LDA+U scheme:
\begin{eqnarray}
H &=&H_{t}+H_{U}  \nonumber \\
H_{t} &=&\sum\limits_{\lambda \lambda ^{\prime }\sigma }t_{\lambda \lambda
^{\prime }}c_{\lambda \sigma }^{+}c_{\lambda ^{\prime }\sigma }  \nonumber \\
H_{U} &=&\frac{1}{2}\sum\limits_{\left\{ \lambda _{i}\right\} \sigma \sigma
^{\prime }}\left\langle \lambda _{1}\lambda _{2}\left| v\right| \lambda
_{1}^{\prime }\lambda _{2}^{\prime }\right\rangle c_{\lambda _{1}\sigma
}^{+}c_{\lambda _{2}\sigma ^{\prime }}^{+}c_{\lambda _{2}^{\prime }\sigma
^{\prime }}c_{\lambda _{1}^{\prime }\sigma \,,}  \label{hamilU}
\end{eqnarray}
where $\lambda =im$ are the site number $\left( i\right) $ and orbital $%
\left( m\right) $ quantum numbers, $\sigma =\uparrow ,\downarrow $ is the
spin projection, $c^{+},c$ are the Fermion creation and annihilation
operators, $H_{t}$ is the effective single-particle Hamiltonian from the
LDA, corrected for the double-counting of average interactions among
correlated electrons as it will be described below. The matrix
elements of the screened Coulomb potential are defined in the standard way 
\begin{equation}
\left\langle 12\left| v\right| 34\right\rangle =\int d{\bf r}d{\bf r}%
^{\prime }\psi _{1}^{\ast }({\bf r})\psi _{2}^{\ast }({\bf r}^{\prime
})v\left( {\bf r-r}^{\prime }\right) \psi _{3}({\bf r})\psi _{4}({\bf r}%
^{\prime }),  \label{coulomb}
\end{equation}
where we define for briefness $\lambda _{1}\equiv 1$ etc. A general SPTF
scheme has been presented recently \cite{KL02}. For $d$
electrons in cubic structures where the one-site Green function is diagonal
in orbital indices the general formalism can be simplified. First, the basic
equation for the $T$-matrix which replaces the effective potential in the
SPTF approach reads  
\begin{eqnarray}
\left\langle 13\left| T^{\sigma \sigma ^{\prime }}\left( i\Omega \right)
\right| 24\right\rangle =\left\langle 13\left| v\right| 24\right\rangle 
\frac{1}{\beta }\sum\limits_{\omega }\sum\limits_{56}\left\langle 13\left|
v\right| 56\right\rangle \times   \nonumber \\
G_{5}^{\sigma }\left( i\omega \right) G_{6}^{\sigma ^{\prime }}\left(
i\Omega -i\omega \right) \left\langle 56\left| T^{\sigma \sigma ^{\prime
}}\left( i\Omega \right) \right| 24\right\rangle ,  \label{tmatrix}
\end{eqnarray}
where $\omega =(2n+1)\pi T$ are the Matsubara frequencies for
temperature $T\equiv\beta^{-1}\ \ $ ($n=0,\pm 1,...$). 

At first, we should take into account the ``Hartree'' and ``Fock'' diagrams
with the replacement of the bare interaction by the $T$-matrix 
\begin{eqnarray}
\Sigma _{12,\sigma }^{\left( TH\right) }\left( i\omega \right)  &=&\frac{1}{%
\beta }\sum\limits_{\Omega }\sum\limits_{3\sigma ^{\prime }}\left\langle
13\left| T^{\sigma \sigma ^{\prime }}\left( i\Omega \right) \right|
23\right\rangle G_{3}^{\sigma ^{\prime }}\left( i\Omega -i\omega \right)  
\nonumber  \\
\Sigma _{12,\sigma }^{\left( TF\right) }\left( i\omega \right)  &=&-\frac{1}{%
\beta }\sum\limits_{\Omega }\sum\limits_{3}\left\langle 13\left| T^{\sigma
\sigma }\left( i\Omega \right) \right| 32\right\rangle G_{3}^{\sigma }\left(
i\Omega -i\omega \right) \, .   \nonumber \\
&& \label{HarFock}
\end{eqnarray}

Now we rewrite the effective Hamiltonian (\ref{hamilU}) with the replacement 
$\left\langle 12\left| v\right| 34\right\rangle $ by $\left\langle 12\left|
T^{\sigma \sigma ^{\prime }}\right| 34\right\rangle $ in $H_{U}$. To
consider the correlation effects described due to P-H channel we have to separate
density ($d$) and magnetic ($m$) channels as in Ref.\cite{BS89}

\begin{eqnarray}
d_{12} &=&\frac{1}{\sqrt{2}}\left( c_{1\uparrow }^{+}c_{2\uparrow
}+c_{1\downarrow }^{+}c_{2\downarrow }\right)   \nonumber \\
m_{12}^{0} &=&\frac{1}{\sqrt{2}}\left( c_{1\uparrow }^{+}c_{2\uparrow
}-c_{1\downarrow }^{+}c_{2\downarrow }\right)   \nonumber \\
m_{12}^{+} &=&c_{1\uparrow }^{+}c_{2\downarrow }  \nonumber \\
m_{12}^{-} &=&c_{1\downarrow }^{+}c_{2\uparrow }\, .  \label{chan}
\end{eqnarray}
Then the interaction Hamiltonian can be rewritten in the following matrix
form 
\begin{equation}
H_{U}=\frac{1}{2}Tr\left( D^{+}\ast V^{\parallel }\ast D+m^{+}\ast
V_{m}^{\perp }\ast m^{-}+m^{-}\ast V_{m}^{\perp }\ast m^{+}\right) , 
\label{hamnew}
\end{equation}
where $\ast $ means the matrix multiplication with respect to the pairs of
orbital indices, e.g. 
\[
\left( V_{m}^{\perp }\ast m^{+}\right) _{11^{\prime
}}=\sum\limits_{34}\left( V_{m}^{\perp }\right) _{11^{\prime },22^{\prime
}}m_{22^{\prime }}^{+}\ .
\]
The supervector D is defined as 
\[
D=\left( d,m^{0}\right) ,D^{+}=\left( 
\begin{array}{c}
d^{+} \\ 
m_{0}^{+}
\end{array}
\right) ,
\]
and the effective interactions have the following form: 
\begin{eqnarray}
&&\left( V_{m}^{\perp }\right) _{11^{\prime },22^{\prime }}=-\left\langle
12\left| T^{\uparrow \downarrow }\right| 2^{\prime }1^{\prime }\right\rangle 
\nonumber \\
&&V^{\parallel }=\left( 
\begin{array}{cc}
V^{dd} & V^{dm} \\ 
V^{md} & V^{dd}
\end{array}
\right)   \nonumber \\
&&V_{11^{\prime },22^{\prime }}^{dd}=\frac{1}{2}\sum\limits_{\sigma \sigma
^{\prime }}\left\langle 12\left| T^{\sigma \sigma ^{\prime }}\right|
1^{\prime }2^{\prime }\right\rangle -\frac{1}{2}\sum\limits_{\sigma
}\left\langle 12\left| T^{\sigma \sigma }\right| 2^{\prime }1^{\prime
}\right\rangle   \nonumber \\
&&V_{11^{\prime },22^{\prime }}^{mm}=\frac{1}{2}\sum\limits_{\sigma \sigma
^{\prime }}\sigma \sigma ^{\prime }\left\langle 12\left| T^{\sigma \sigma
^{\prime }}\right| 1^{\prime }2^{\prime }\right\rangle -\frac{1}{2}%
\sum\limits_{\sigma }\left\langle 12\left| T^{\sigma \sigma }\right|
2^{\prime }1^{\prime }\right\rangle   \nonumber
\end{eqnarray}
\begin{eqnarray}
V_{11^{\prime },22^{\prime }}^{dm}=V_{22^{\prime },11^{\prime }}^{md}=\hspace{7cm} 
\nonumber \\
\hspace{2cm} \frac{1}{2}[\left\langle 12\left| T^{\uparrow \uparrow }\right| 1^{\prime
}2^{\prime }\right\rangle -\left\langle 12\left| T^{\downarrow \downarrow
}\right| 1^{\prime }2^{\prime }\right\rangle -\left\langle 12\left|
T^{\uparrow \downarrow }\right| 1^{\prime }2^{\prime }\right\rangle 
\nonumber \\
+\left\langle 12\left| T^{\downarrow \uparrow }\right| 1^{\prime }2^{\prime
}\right\rangle -\left\langle 12\left| T^{\uparrow \uparrow }\right|
2^{\prime }1^{\prime }\right\rangle +\left\langle 12\left| T^{\downarrow
\downarrow }\right| 2^{\prime }1^{\prime }\right\rangle ]\ .  \label{effpotent}
\end{eqnarray}
To calculate the particle-hole (P-H) contribution to the electron
self-energy we first have to write the expressions for the generalized
susceptibilities, both transverse $\chi ^{\perp }$ and longitudinal $\chi
^{\parallel }$. One has 
\begin{equation}
\chi ^{+-}(i\omega )=\left[ 1+V_{m}^{\perp }\ast \Gamma ^{\uparrow
\downarrow }(i\omega )\right] ^{-1}\ast \Gamma ^{\uparrow \downarrow
}(i\omega )\,,  \label{xi+-}
\end{equation}
where 
\begin{equation}
\Gamma _{12,34}^{\sigma \sigma ^{\prime }}\left( \tau \right)
=-G_{2}^{\sigma }\left( \tau \right) G_{1}^{\sigma ^{\prime }}\left( -\tau
\right) \delta _{23}\delta _{14}  \label{gamma}
\end{equation}
is an ``empty loop'' susceptibility and $\Gamma (i\omega )$ is its Fourier
transform, $\tau $ is the imaginary time. The corresponding longitudinal
susceptibility matrix has a more complicated form: 
\begin{equation}
\chi ^{\parallel }(i\omega )=\left[ 1+V^{\parallel }\ast \chi
_{0}^{\parallel }(i\omega )\right] ^{-1}\ast \chi _{0}^{\parallel }(i\omega
),  \label{xipar}
\end{equation}
and the matrix of the bare longitudinal susceptibility is 
\begin{equation}
\chi _{0}^{\parallel }=\frac{1}{2}\left( 
\begin{array}{cc}
\Gamma ^{\uparrow \uparrow }+\Gamma ^{\downarrow \downarrow }\, & \,\Gamma
^{\uparrow \uparrow }-\Gamma ^{\downarrow \downarrow } \\ 
\Gamma ^{\uparrow \uparrow }-\Gamma ^{\downarrow \downarrow }\, & \,\Gamma
^{\uparrow \uparrow }+\Gamma ^{\downarrow \downarrow }
\end{array}
\right) ,  \label{xi0par}
\end{equation}
in the $dd$-, $dm^{0}$-, $m^{0}d$-, and $m^{0}m^{0}$- channels ($d,m^{0}=1,2$
in the supermatrix indices). An important feature of these equations is the
coupling of longitudinal magnetic fluctuations and of density fluctuations.
It is not present in the one-band Hubbard model due to the absence of the
interaction of electrons with parallel spins. For this case
Eqs. (\ref{xi+-}) and (\ref{xipar}) coincide with the well-known result
of Izuyama $et.\ al.$ \cite{IKK63}.

Now we can write the particle-hole contribution to the self-energy. Similar
to Ref.\cite{KL99} one has 
\begin{equation}
\Sigma _{12,\sigma }^{(ph)}\left( \tau \right) =\sum\limits_{34,\sigma
^{\prime }}W_{13,42}^{\sigma \sigma ^{\prime }}\left( \tau \right)
G_{34}^{\sigma ^{\prime }}\left( \tau \right) ,  \label{sigph}
\end{equation}
with the P-H fluctuation potential matrix:

\begin{equation}
W^{\sigma \sigma ^{\prime }}\left( i\omega \right) =\left[ 
\begin{array}{cc}
W^{\uparrow \uparrow }\left( i\omega \right) & W^{\perp }\left( i\omega
\right) \\ 
W^{\perp }\left( i\omega \right) & W^{\downarrow \downarrow }\left( i\omega
\right)
\end{array}
\right] ,  \label{wpp}
\end{equation}
were the spin-dependent effective potentials are defined as 
\begin{eqnarray}
W^{\uparrow \uparrow }&=&\frac{1}{2}V^{\parallel }\ast \left[ \chi
^{\parallel }-\chi _{0}^{\parallel }\right] \ast V^{\parallel }  \nonumber \\
W^{\downarrow \downarrow }&=&\frac{1}{2}V^{\parallel }\ast \left[ \widetilde{%
\chi }^{\parallel }-\widetilde{\chi }_{0}^{\parallel }\right] \ast
V^{\parallel }  \nonumber \\
W^{\uparrow \downarrow }&=&V_{m}^{\perp }\ast \left[ \chi ^{+-}-\chi
_{0}^{+-}\right] \ast V_{m}^{\perp }  \nonumber \\
W^{\downarrow \uparrow }&=&V_{m}^{\perp }\ast \left[ \chi ^{-+}-\chi
_{0}^{-+}\right] \ast V_{m}^{\perp }\ .
\end{eqnarray}
Here $\widetilde{\chi }^{\parallel },\widetilde{\chi }_{0}^{\parallel }$
differ from $\chi ^{\parallel },\chi _{0}^{\parallel }$ by the replacement
of $\Gamma ^{\uparrow \uparrow }\Leftrightarrow \Gamma ^{\downarrow
\downarrow }$ in Eq.(\ref{xi0par}). We have subtracted the second-order
contributions since they have already been taken into account in Eq.(\ref
{HarFock}).

Our final expression for the self energy is 
\begin{equation}
\Sigma =\Sigma ^{(TH)}+\Sigma ^{(TF)}+\Sigma ^{(PH)}\ .  \label{sigmatot}
\end{equation}
This formulation takes into account accurately spin-polaron effects because
of the interaction with magnetic fluctuations \cite{IK90,IK94}, the energy
dependence of the $T$-matrix which is important for describing the satellite
effects in Ni \cite{L81}, contains exact second-order terms in $v$ and
is rigorous (because of the first term) for almost filled or almost empty
bands.

Since the LSDA Green's function already contains the average
electron-electron interaction, in Eqs.\ (\ref{HarFock}) and
(\ref{sigph}) the static part of the self-energy $\Sigma^{\sigma}(0)$ is not included, i.e. we have		

\begin{equation}\label{tildeS}
{\tilde\Sigma}^{\sigma}(i\omega)=\Sigma^{\sigma}(i\omega)-\Sigma^{\sigma}(0).
\end{equation}

\begin{figure}[t]
%
%
\begin{center}
\leavevmode
\includegraphics[width=0.55\textwidth,angle=-90,clip]{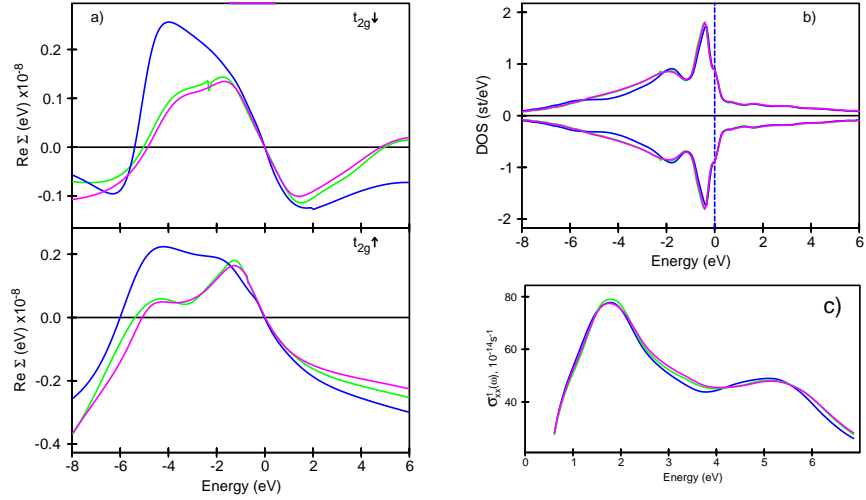}
\end{center}
\caption {\label{temperature} The self-energy (a) of Fe for three different
temperatures and corresponding densities of states (b) and optical
conductivities spectra (c). Full, dashed and dotted lines correspond to
$T=125K$, $T=300K$ and $T=900K$, respectively.}
\end{figure}

\section{Results and discussion}

The matrix elements of $v$ appearing in Eq.(\ref{coulomb}) can be
calculated in terms of two parameters - the averaged screened Coulomb
interaction $U$ and exchange interaction $J$ \cite{KL02}. 
The screening of the exchange interaction is usually small and the
value of $J$ can be calculated directly. Moreover numeric calculations
show that the value of $J$ for all $3d$ elements is practically the
same and approximately equal to 0.9~eV. This value has been adopted for
all our calculations presented here. At the same time  direct Coulomb
interaction undergoes substantional screening and one has to be
extremely careful  making the choice for this parameter. There
are some prescription how one can get it within  constraint LDA
calculation \cite{JG89}. However, results obtained in this way depend
noticeably on the choice of the basis functions, way of accounting for
hybridization etc. Nevertheless the order of magnitude coming out from
various approaches is the same giving the value of $U$  in the
range 1--4~eV.  In the present paper we are discussing the influence of the
choice of $U$ on the calculated optical spectra.

Another parameter entering SPTF equations is temperature. For a moment
we are more interested in the low temperature properties while
computationally the higher the tempreture is, less computationally demanding are
the calculations. This is why we decided first to consider the dependence of the
self-energy on the temperature. 

In Fig. \ref{temperature} we show the self-energy obtained for Fe for three
different temperatures as well as corresponding densities of states and optical
conductivities spectra. One can see that despite the differences in $\Sigma$ are
quite noticeable this leads only to moderate changes in the density of states
and does not affect the optical conductivity.

\begin{figure}[t]
%
%
\begin{center}
\leavevmode
\includegraphics[width=0.46\textwidth,angle=-90,clip]{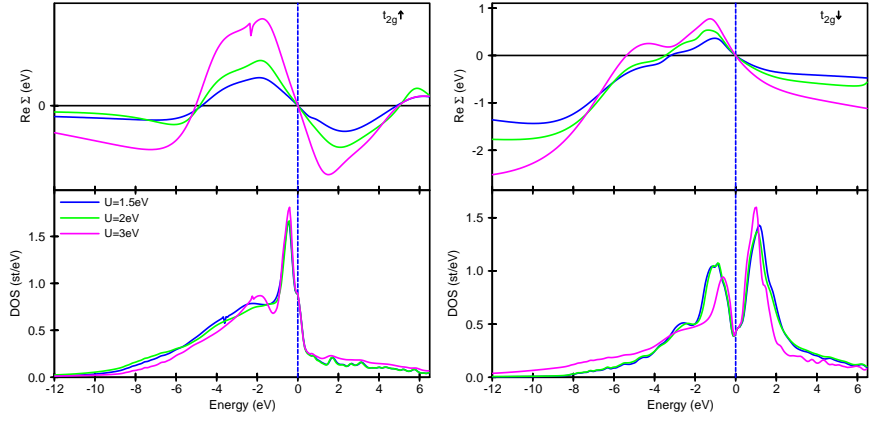}
\end{center}
\caption {\label{fedosfig} The real part of $t_{2g}$ component of
$\Sigma$ for $T=300K$ in Fe for various values of $U$; left: spin-up, right: spin-down.}
\end{figure}

\begin{figure}[b]
%
%
\begin{center}
\leavevmode
\includegraphics[width=0.41\textwidth,angle=-90,clip]{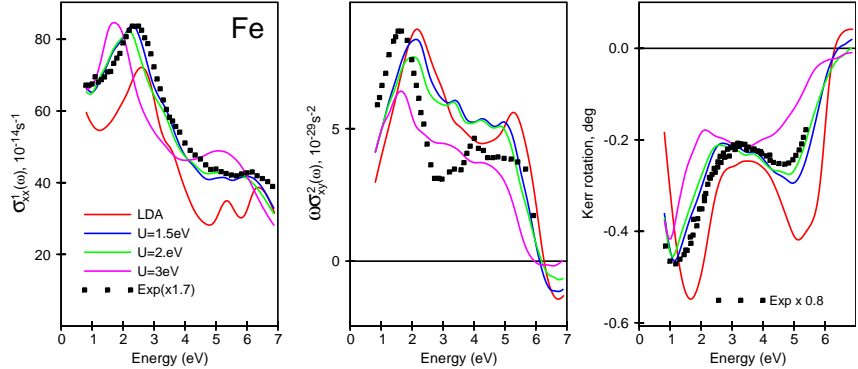}
\end{center}
\caption {\label{feoptfig} Optical conductivity (left: diagonal; middle:
off-diagonal) and polar Kerr rotation (right)
spectra in comparison with the experimental data of Fe. Experimental
data for conductivity are taken from Ref.~\cite{Fe_sig1,Fe_sig2}; Kerr
rotation spectra - from Ref.~\cite{FeNi_kerr}}
\end{figure}

Much more important for the results is the parameter
$U$. Fig.~\ref{fedosfig} shows as an example the real part $t_{2g}$ 
component of $\Sigma$ for $T=300K$ in Fe for various values of $U$. 
Despite the overall shape of the curve is practically the same the
magnitude of the self-energy increasing with increase of $U$ as 
it is expected from the analytical expressions. This
change in self-energy leads to corresponding changes in the densities
of states especially noticeable for the minority spin subband. The
influence of the choice of $U$ on the optical properties is even more
pronounced (see Fig.~\ref{feoptfig}).  The low energy peak in the diagonal part of the
optical conductivity  shifts to the lower energies reaching the
experimental position already for $U$=1.5~eV. In  the high energy part
of the spectra large values of $U$ lead to a structure around 5~eV
not seen in experiment. 

Again, the value $U$=1.5~eV gives also the best description for the
shape of the experimental curve. (Note, that the experimental results
for $\sigma_{xx}^{1}(\omega)$ are multiplied by a factor of 1.7 to make the
comparison more obvious.) A rather different situation occurs for
the off-diagonal part of the optical conductivity. The low energy peak can
be brought to the proper position only with $U$=4~eV, at the same time
the shape of the theoretical curves above 2.5~eV has a rather different structure in
comparison with the experimental one, only crossing the zero axis at the same
energy. However, a direct comparison of calculated $\sigma_{xy}$ data with
experimental ones may be somewhat misleading as experimentally this
quantity cannot be measured directly and is usually obtained from 
ellipsometric measurements and measurements of the Kerr rotation
spectra. Thus in the left panel of the Fig.~\ref{feoptfig} we show our
results for the calculated polar Kerr rotation spectra in comparison with 
experimental data. As one can see again the DMFT calculation with
$U=1.5$~eV describes the experimental data in a rather satisfactory way.

\begin{figure}[t]
%
%
\begin{center}
\leavevmode
\includegraphics[width=0.46\textwidth,angle=-90,clip]{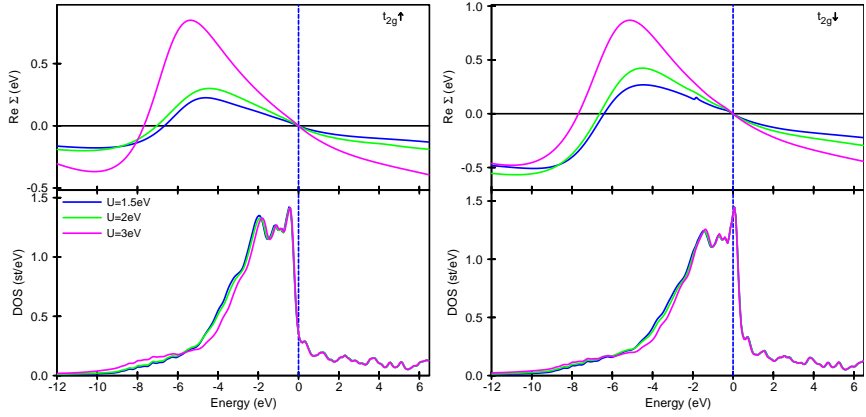}
\end{center}
\caption {\label{nidosfig} The real part of $t_{2g}$ self-energy for $\
U=1.5,2,3$~eV and corresponding DOS plots for Ni; left: spin-up, right: spin-down.}
\end{figure}

If for Fe LSDA calculations already give a reasonable description of the
optical properties and the many-body correlation effects, which
improves only minor details, the situation in Ni is quite
different. It is well-known that LSDA fails to describe the 
bandwidth for Ni, causing problems in the theoretical interpretation
of all the spectroscopic experiments such as photoemission, x-ray
emission, optics, etc. The main reason for this is the underestimation of
electron-electron correlations which appear to be relatively strong in this
metal. Again, as in the case of Fe, we carried out calculations with
different values of $U$ to find the best description of the spectral
properties of Ni. In Fig.~\ref{nidosfig} we show the real part of the $t_{2g}$ self-energy
for $U = 1.5, 2, 3$~eV as well as corresponding DOS plots. 

\begin{figure}[t]
%
%
\begin{center}
\leavevmode
\includegraphics[width=0.42\textwidth,angle=-90,clip]{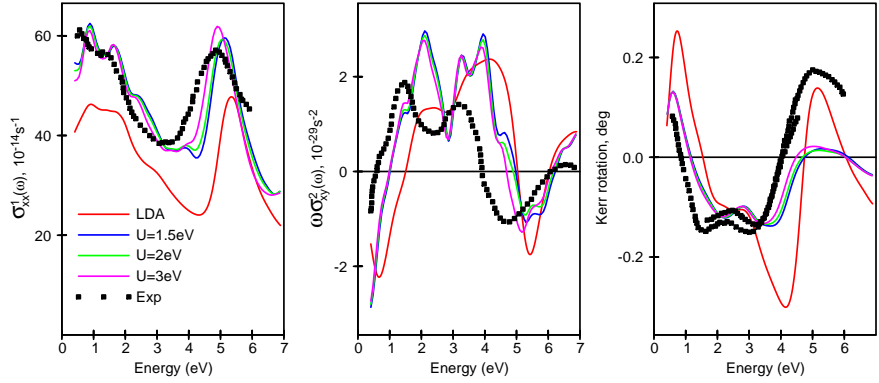}
\end{center}
\caption {\label{nioptfig} Optical conductivity and polar Kerr rotation
spectra in comparison with the experimental data of Ni. Experimental
data for conductivity are taken from Ref.~\cite{Ni_sig1,Ni_sig2}. Kerr
rotation spectra - from Ref.~\cite{FeNi_kerr}}
\end{figure}

Despite the changes in the amplitude of the self-energy are important, all self-energies
lead to rather small changes in the density of states, narrowing
somewhat the bandwidth only and developing a low energy tail. Nevertheless
the diagonal part of the conductivity which reflects the convolution of the
occupied and unoccupied states is much more affected by the choice of
$U$. The main change can be seen in the position of high energy peak
which is placed by LDA about 1~eV higher in comparison with
experiment. Accounting for the correlation effects shifts this maximum
bringing it to the proper position for $U=3$~eV. The low energy part of
the spectra does not reflect too much influence of the $U$ parameter and
deviates just slightly from the experimental curve. For the
off-diagonal part of the conductivity an improvement as compared to 
LDA is not so pronounced as for the diagonal one, though the spectra getting closer
to experiment. It is worth to note that the actual value of $U$ doesn't
change  the calculated spectrum of $\omega\sigma^2_{xy}(\omega)$. But again,
as mentioned in the case of Fe 
it is worth to compare calculations with directly measured Kerr rotation
spectra presented in Fig.~\ref{nioptfig}. As one can see, the improvement
compared to LSDA results is substantional
but our results are still far from experiment concerning the peak position both
in the infrared and visible parts of the spectra. This disagreement is
apparently coming from the approximation that has been made and is much more
pronounced in the off-diagonal part of conductivity as it is more
sensitive to the details of the electronic structure being the result of
complex interplay of exchange splitting and spin-orbit coupling. 

It is still unclear whether the mentioned problems are coming from the single-site
approximation for the self-energy (DMFT) itself or whether they are reflecting the limitations of
the simplified FLEX method of solving the impurity many-body problem. To
find out an answer more elaborated solvers like QMC have to be used.

\section{Conclusion and outlook}
In the present paper we show a way to account for the particle-particle
correlations in the theoretical description of optical and magneto-optical
properties of the ferromagnetic 3$d$ metals. We show that the dynamical
correlations play an important role even in weakly correlated
materials like Fe and can substantially change the shape of the spectra for
moderately correlated Ni. Even a rather simple way of accounting for
dynamic correlation allows to improve theoretical results
substantionally though not giving the perfect agreement with experiment. 

Thus to go further one has to use more elaborated technique to obtain the
self-energy both within DMFT and beyond (for example, new DMFT+GW
approximation). Work along this line is in progress.

\end{document}